\documentclass[twocolumn, tighten,  times, 12pt,floatfix]{aastex631}

\usepackage{amsmath}
\usepackage{ulem}
\usepackage{color}
\usepackage{appendix}
\usepackage{float}
\usepackage{microtype}

\shorttitle{AASTeX v6.3.1 Sample article}
\shortauthors{Meng et al.}

\graphicspath{{./}{figures/}}

\begin{document}
\begin{sloppypar}

\title{The Relativistic Spin Precession in the Compact Double Neutron Star System PSR~J1946+2052}

\author{Lingqi Meng}
\affiliation{National Astronomical Observatories, Chinese Academy of Sciences, Beijing 100101, China}
% \affiliation{College of Astronomy and Space Sciences, University of Chinese Academy of Sciences, Beijing 100049, China}
\affiliation{School of Astronomy and Space Science, University of Chinese Academy of Sciences, Beijing 100049, China}
\author{Weiwei Zhu}
\affiliation{National Astronomical Observatories, Chinese Academy of Sciences, Beijing 100101, China}
\affiliation{Institute for Frontiers in Astronomy and Astrophysics, Beijing Normal University, Beijing 102206, China}
\author{Michael Kramer}
\affiliation{Max-Planck-Institut für Radioastronomie, Auf dem Hügel 69, D-53121 Bonn, Germany}
\affiliation{Jodrell Bank Centre for Astrophysics, Department of Physics and Astronomy, University of Manchester, M13 9PL Manchester, UK}
\author{Xueli Miao}
\affiliation{National Astronomical Observatories, Chinese Academy of Sciences, Beijing 100101, China}
\author{Gregory Desvignes}
\affiliation{Max-Planck-Institut für Radioastronomie, Auf dem Hügel 69, D-53121 Bonn, Germany}
\author{Lijing Shao}
\affiliation{Kavli Institute for Astronomy and Astrophysics, Peking University, Beijing 100871, China}
\affiliation{National Astronomical Observatories, Chinese Academy of Sciences, Beijing 100101, China}
\affiliation{Max-Planck-Institut für Radioastronomie, Auf dem Hügel 69, D-53121 Bonn, Germany}
\author{Huanchen Hu}
\affiliation{Max-Planck-Institut für Radioastronomie, Auf dem Hügel 69, D-53121 Bonn, Germany}
\author{Paulo~C.~C.~Freire}
\affiliation{Max-Planck-Institut für Radioastronomie, Auf dem Hügel 69, D-53121 Bonn, Germany}
\author{Yongkun Zhang}
\affiliation{National Astronomical Observatories, Chinese Academy of Sciences, Beijing 100101, China}
\affiliation{School of Astronomy and Space Science, University of Chinese Academy of Sciences, Beijing 100049, China}
\author{Mengyao Xue}
\affiliation{National Astronomical Observatories, Chinese Academy of Sciences, Beijing 100101, China}
\author{Ziyao Fang}
\affiliation{National Astronomical Observatories, Chinese Academy of Sciences, Beijing 100101, China}
\affiliation{The Key Laboratory of Cosmic Rays (Tibet University), Ministry of Education, Lhasa 850000, China}
\author{David J. Champion}
\affiliation{Max-Planck-Institut für Radioastronomie, Auf dem Hügel 69, D-53121 Bonn, Germany}
\author{Mao Yuan}
\affiliation{National Space Science Center, Chinese Academy of Sciences, Beijing 100190, China}
\author{Chenchen Miao}
\affiliation{Research Institute of Artificial Intelligence, Zhejiang Lab, Hangzhou, Zhejiang 311121, China}
\author{Jiarui Niu}
\affiliation{National Astronomical Observatories, Chinese Academy of Sciences, Beijing 100101, China}
\affiliation{School of Astronomy and Space Science, University of Chinese Academy of Sciences, Beijing 100049, China}
\author{Qiuyang Fu}
\affiliation{National Astronomical Observatories, Chinese Academy of Sciences, Beijing 100101, China}
\affiliation{School of Astronomy and Space Science, University of Chinese Academy of Sciences, Beijing 100049, China}
\author{Jumei Yao}
\affiliation{Xinjiang Astronomical Observatory, Chinese Academy of Sciences, 150, Science 1-Street, Urumqi, Xinjiang 830011, China}
\author{Yanjun Guo}
\affiliation{Max-Planck-Institut für Radioastronomie, Auf dem Hügel 69, D-53121 Bonn, Germany}
\author{Chengmin Zhang}
\affiliation{National Astronomical Observatories, Chinese Academy of Sciences, Beijing 100101, China}

\begin{abstract}
We observe systematic profile changes in the visible pulsar of the compact double neutron star system PSR~J1946+2052 
using observations with the Five-hundred-meter Aperture Spherical radio Telescope (FAST). 
The interpulse of PSR~J1946+2052 changed from single-peak to double-peak shape from 2018 to 2021. 
We attribute this evolution as the result of the relativistic spin precession of the pulsar. 
With the high sensitivity of FAST, we also measure significant polarization for the first time, allowing us to
model this with the precessional rotating vector model. Assuming, to the first order, a circular 
hollow-cone-like emission beam pattern and taking the validity of general relativity, we derive the binary's orbital inclination angle (${63^\circ}^{+5^\circ}_{-3^\circ}$) and pulsar's spin geometry.
Pulsar's spin vector and the orbital angular momentum vector are found to be only slightly misaligned (${0.21^\circ}^{+0.28^\circ}_{-0.10^\circ}$).The quoted uncertainties do not reflect the systematic uncertainties introduced by our model assumptions. 
By simulating future observations of profile and polarization evolution, we estimate that we could constrain the precession rate within a $43\%$ uncertainty in 9 years. Hence, we suggest that the system's profile evolution could be combined with precise pulsar timing to test general relativity in the future. 

\end{abstract}


\section{Introduction} \label{sec:intro}
The discovery of pulsars in double neutron star (DNS) systems \citep[e.g. PSR B1913+16 ][]{hulse1975discovery} unlocks ideal laboratories for testing theories of gravitation. 
Relativistic effects, such as Shapiro delay, Einstein delay, orbital period
decay caused by gravitational wave emission, and periastron advance, can be observed in pulsar timing through the measurements of the post-Keplerian parameters of pulsar binary systems \citep{damour1985general,damour1986general}.
Timing of PSR B1913+16 and PSR J0737$-$3039A/B has already stringently tested general relativity (GR), like \cite{2005ASPC..328...25W,2016ApJ...829...55W,kramer2006tests,kramer2021strong}.

Relativistic spin precession is another relativistic effect caused by the curvature of space-time in the presence of a large mass object.
This precession causes the pulsar's spin vector to precess around the total angular momentum vector and moves the 
pulsar's emission beam with respect to our line of sight.
The relativistic spin precession is usually not directly detectable through pulsar timing,
%This relativistic precession cannot be detected by pulsar timing directly, 
but sometimes can be observed through the temporal evolution of the pulse profile and its polarization.

The first indication for profile changes caused by relativistic spin precession was given by \cite{weisberg1989evidence},
observing the relative amplitude of the two prominent components of PSR~B1913+16's profile changing over time. 
A systematic decrease in the pulse width change was later detected, and the profile changes were analyzed and modelled by \cite{kramer1998determination}. A quantitative test of the precession rate was not possible, but the changes were found to be 
consistent with the expectation from GR.
After PSR~B1913+16, the relativistic spin precession has been detected through profile change in more pulsar binary systems, 
such as PSR J1537+1155 \citep{stairs2004measurement}, PSR~J1141$-$6545 \citep{2005ApJ...624..906H}, PSR~J0737$-$3039B \citep{2005ApJ...624L.113B}, PSR J1906+0746 \citep{2006ApJ...640..428L,desvignes2019radio} and, more recently,
in PSR J1757$-$1854 \citep{2023MNRAS.523.5064C}.
Some pulsars in DNSs, such as PSR J0737$-$3039A, show no sign of temporal 
profile evolution. This could be due to a small misalignment angle between the pulsar's spin vector and the orbital angular momentum, as inferred for PSR J0737$-$3039A \citep{ferdman2013double}. As the  polarization position angle (PPA) swing is a marker
of pulsar geometry, the temporal evolution of the PPA is also a potential (and powerful) 
indicator of relativistic spin precession \citep{stairs2004measurement,desvignes2019radio}.

Assuming GR, the rate of spin precession of the pulsar is determined from the system's Keplerian parameters and two objects' masses
\citep{bo75}:
\begin{equation}
\Omega_{\rm SO}= \frac{1}{2}\left (\frac{GM_{\odot}}{c^3}\right )^{2/3}\left(\frac{P_b}{2\pi}\right)^{-5/3}\frac{m_2(4m_1+3m_2)}{(1-e^2)(m_1+m_2)^{4/3}},
\label{eq:omdot}
\end{equation}
where $G$ is the gravitational constant, $c$ is the velocity of light, $P_b$ is the orbital period, $e$ is the orbital eccentricity, $m_1$ and $m_2$ are masses of the pulsar and companion measured in units of solar mass. 
The Keplerian parameters such as $P_b$ and $e$ can be derived through pulsar timing, and masses can be measured if more than two post-Keplerian parameters are detected. 
If the spin precession rate is derived independently through the profile and PPA evolution, then it can be used to test GR \citep{stairs2004measurement,2014ApJ...787...82F,desvignes2019radio}.
In the case of PSR~J1906+0746, the measured spin precession rate agrees with GR to within 5\% \citep{desvignes2019radio}.
In the particular case of PSR~J0737$-$3039A/B, the spin precession rate is determined through modelling pulsar B's eclipse morphology evolution and that leads to a rate in agreement with GR to within 13\% in \cite{breton2008relativistic} and recently 6\% in \cite{2024A&A...682A..26L}.
At last, if the spin precession rate could not be directly measured, one still can use the relativistic spin precession to model the binary system's geometry by assuming GR is correct \citep{kramer1998determination,2010ApJ...710.1694M,ferdman2013double}.

\cite{stovall2018palfa} discovered the DNS system PSR~J1946+2052, which has the shortest orbital period among known DNS systems at 1.88~h. 
The eccentricity of the orbit is 0.064.
They detected only one post-Keplerian parameter, the periastron advance, to be $25.6^\circ\ {\rm yr}^{-1}$, indicating the system's total mass is $2.50(4)~ \rm{M_\odot}$ under the validity of GR.
The pulsar was considered to be the primary neutron star of the system due to the spin period of 16.9~ms and the magnetic field strength at the surface of $4 \times 10^9$~G.  
They suggested that this system resembles the double pulsar system PSR~J0737-3039A/B in configuration and binary evolution. 
Therefore, they proposed that the pulsar's spin vector could be nearly aligned with the orbital angular momentum vector and thus spin precession should be hard to observe. 
In their work, they did not detect any polarization.

The Five-hundred-meter Aperture Spherical radio Telescope (FAST) is a single-dish radio telescope \citep{nan2008introduction,nan2011five} and is operating since 2019 \citep{jiang2020fundamental}. Owing to the high sensitivity of FAST, we report here the detection of polarization and significant profile changes for
PSR~J1946+2052 for the first time. In section \ref{sec:obs}, we describe the data reduction and specifically demonstrate the removal of the ionospheric Faraday rotation effect.
In section \ref{sec:results}, we show the results of our analysis of pulse profile and polarization and model the geometry of this pulsar system. 
In section \ref{sec:discussion} we discuss the evolution of this DNS and predict how the measurement of spin precession rate would advance in future observations. 

\begin{figure*}[h]
  \begin{center}
    \includegraphics[width= 2\columnwidth]{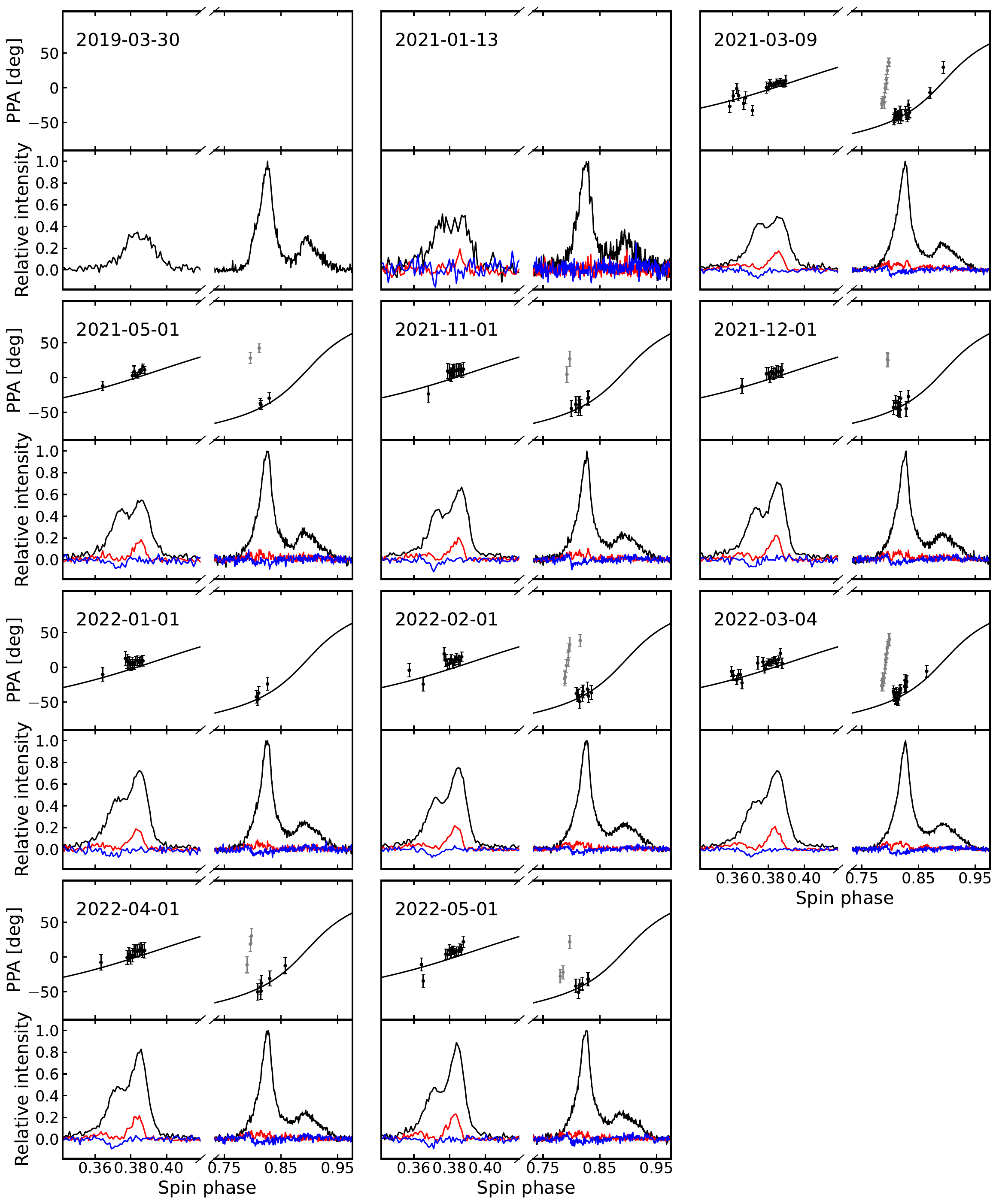}
\end{center}
\caption{This figure presents the pulse profile and polarization information. In the bottom panels, we present the profile of each observation's relative intensity, linear polarization, and circular polarization with black, red, and blue lines respectively. PPAs and precessional RVM curves are shown in the top panels. Data from 2019-03-30 do not have polarization information and position angle because they are formed by 2 polarization. The grey points indicate the part of PPA that we consider to have orthogonal polarization mode. \label{fig:profiles}}
\end{figure*}

\section{Observation and data processing}\label{sec:obs}
We observed PSR~J1946+2052 with the centre beam of FAST's 19-beam receiver at a central frequency of 1250MHz using a bandwidth of 500MHz for 11 epochs.
The first session, conducted on March 30, 2019, was observed in 2 polarization (AA, BB) mode, while the remaining sessions were carried out in full Stokes mode in 2021 and 2022.
Details of the length and date of observations are shown in Table \ref{tab:obslist}. 
We recorded all data in PSRFITS \citep{hanisch2001definition} format with a sampling time of 49.152~$\mu s$ and 8-bit digitization. 
To calibrate each observation's polarization, we pointed the telescope off the pulsar and turned on the noise diode in a 0.2-s cycle for 1 minute before observing the pulsar.
\begin{table}[ht]
\centering
\caption{Date, length, and ionosphere RM of each observation}
\label{tab:obslist}
 \begin{tabular}{c c c} 
 \hline
 Date & Length of observation & Ionospheric RM\\ 
      & (s)  &  (rad~m$^{-2}$) \\[0.5ex]
 \hline
 2019-03-30 & 2400 & -\\ 
 2021-01-13 & 1800 & -\\ 
 2021-03-09 & 7215 & 0.98(3)\\
 2021-05-01 & 2910 & 0.43(2)\\
 2021-11-01 & 3015 & 2.66(10)\\
 2021-12-01 & 3600 & 2.20(2)\\
 2022-01-01 & 7200 & 1.88(4)\\
 2022-02-01 & 7200 & 1.95(5)\\
 2022-03-04 & 7200 & 1.34(7)\\
 2022-04-01 & 7200 & 2.59(8)\\
 2022-05-01 & 7200 & 1.69(16)\\
 %\hline
 \hline
\end{tabular}
\tablecomments {``-'' indicates 2-pol or low SNR observations where no polarization information was available. }
\end{table}

We used the \texttt{DSPSR} \citep{van2011dspsr} analysis program and the \texttt{PSRCHIVE} \citep{hotan2004psrchive} software package to fold and de-disperse the pulsar data, to eliminate radio frequency interference (RFI), and to calibrate the polarization.
We obtained the ephemeris from \texttt{PSRCAT}\footnote{\url{https://www.atnf.csiro.au/research/pulsar/psrcat/}}\citep{2005AJ....129.1993M} and updated it with our own time-of-arrival (TOA) measurements using \texttt{TEMPO}\footnote{\url{http://tempo.sourceforge.net}} \citep{2015ascl.soft09002N}. 
With the local solution of the ephemeris derived from \texttt{TEMPO}, for each epoch, we folded the data 
and scrunched time.
To avoid effects due to uncertainties in the long-term ephemeris, we 
aligned pulses from different observations by using the maximum of the normalized pulse profile.
The Rotation Measure ($\rm RM$) was measured using \texttt{rmfit} in \texttt{PSRCHIVE}.
However, due to the low Signal-to-Noise Ratio (SNR) and low degree of linear polarization, we could not derive a precise $\rm RM$ 
for the whole set of observations.
Instead, we used the observation with the highest SNR to obtain $\rm RM$=$-27(11)~\rm{rad~m}^{-2}$ and removed the Faraday rotation effect in polarization with a fixed RM $-27~$rad~m$^{-2}$ to analyze the relative change of the geometrical parameters (see Sec. \ref{sec:results}).
Then we computed the Stokes parameters I, Q, U, and V and exported the archive data into a {\it{numpy}} \citep{2007CSE.....9c..21P} readable format using the Python package of \texttt{PSRCHIVE}.

% \textbf{Polarization is sensitive to $RM$.}
RM is a crucial parameter in studying pulsar polarization. 
For example, $1~$rad~$\rm{m}^{-2}$ shifts the PPA by almost $14^\circ$ at 1250~MHz.
As explained in \cite{sotomayor2013calibrating}:
\begin{equation}
\rm {RM = RM_{\rm int}+RM_{\rm ion}+RM_{\rm ism}+RM_{\rm igm}},
\label{eq:RM}
\end{equation}
where $\rm {RM_{\rm {int}}, RM_{\rm ion}, RM_{\rm ism}, RM_{\rm igm}}$ denote the intrinsic RM and RM of the ionosphere, interstellar medium, inter-galactic medium respectively, where the latter is not applicable in this case. 
We can also expect a change 
in $\rm {RM_{ism}}$ to be small, based on typical line-of-sights with a relatively uniform magnetic field along
those and conservative estimates of a $\delta\rm{DM}$ of approximately $10^{-4}$ pc cm$^{-3}$.
Hence,
we consider mainly the influence of ionospheric rotation measure ($\rm{RM}_{\rm ion}$) 
and its variation on the measurement of the PPA.

\begin{figure}[htbp]
\vspace{0.3cm}
\begin{center}
    \includegraphics[width= \columnwidth]{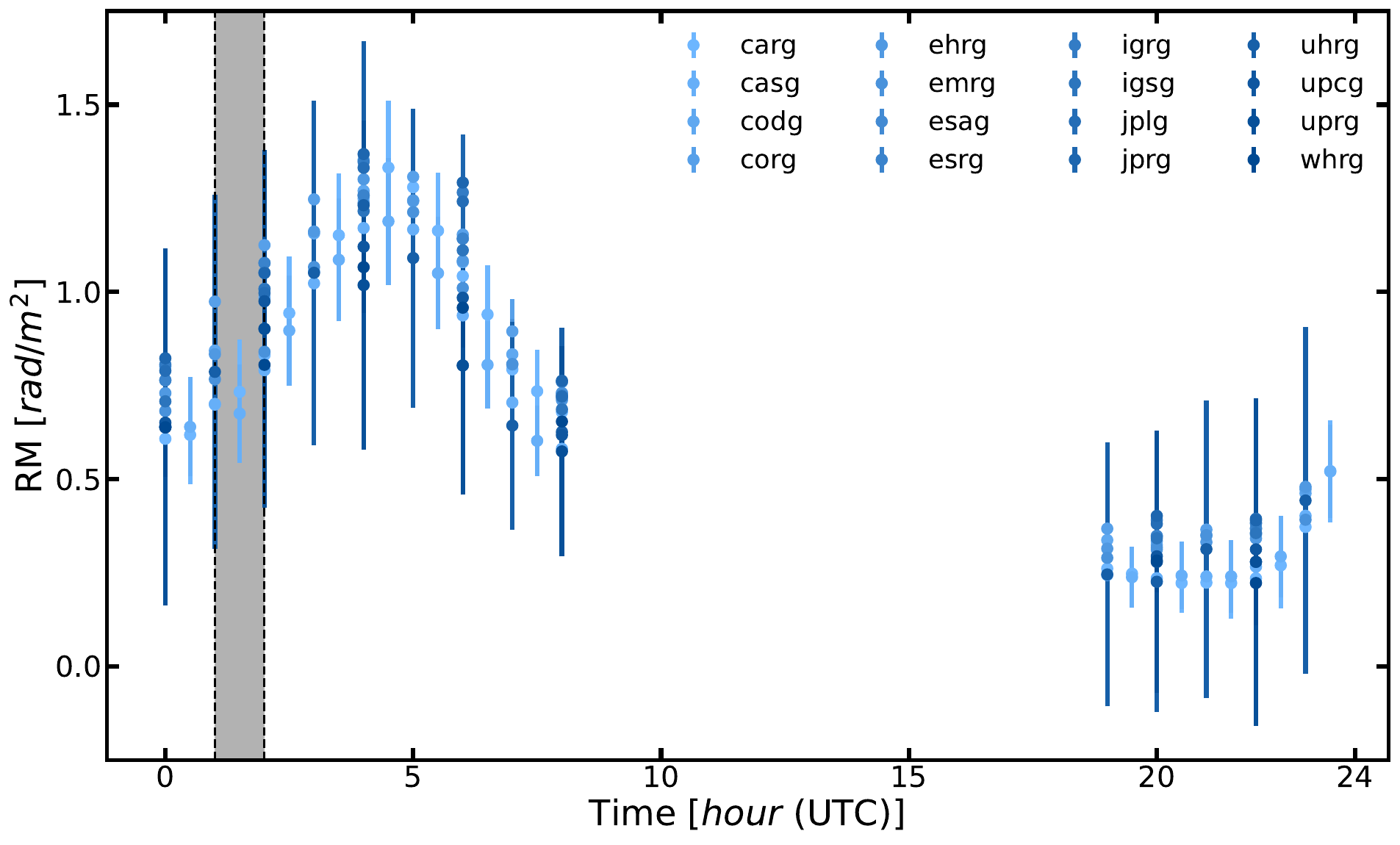}
\end{center}
\vspace{-0.5cm}
\caption{${\rm RM_{ion}}$ values on 2021-03-09 derived from different GIM analysis centres are indicated by various colours.
The blank parts without points indicate the time when PSR~J1946+2052 is not visible to FAST. The grey area represents the time span of our observation.\label{fig:ionrm}}
% \vspace{-0.3cm}
\end{figure}

We calculated $\rm{RM}_{\rm ion}$ of each epoch by using the {\it{ionFR}} package \citep{sotomayor2013calibrating} and global ionospheric map (GIM) products. The GIM products were generated by different analysis centres using total electron content (TEC) gathered from global positioning system (GPS) stations around the world, combined with a specific mathematical model \citep{li2017evaluation}. 
For each observation, dozens of GIM analysis centres can provide measurements of the ionospheric RM, and their estimates may differ slightly due to model dependence.
Figure \ref{fig:ionrm} shows an example of the $\rm{RM}_{\rm ion}$ of PSR J1946+2052 derived from GIM products on March 9, 2021, which are available on NASA's website.\footnote{\url{cddis.nasa.gov/archive/gnss/products/ionex/}} 
To remove the bias in the GIM products, we interpolated values of $\rm{RM}_{\rm ion}$ and their errors during the time span of our observations and obtained the weighted average and error for each analysis centre. 
Then we weighted these $\rm{RM}_{\rm ion}$ from different analysis centres with their errors and derived the average and the uncertainty of each day’s ionospheric RM. 
The $\rm {RM_{ion}}$ values for each observation epoch are presented in Table \ref{tab:obslist}.

We have detected polarization in observations with FAST lasting longer than one hour. The profiles at 1.25 GHz are presented in Fig. \ref{fig:profiles} with 1024 phase bins. We consider the emission feature near pulse phases $\sim 0.36 - 0.40$ to be the interpulse, complementing the stronger main pulse at pulse phases $\sim 0.77 - 0.93$.
% The PPAs are derived from $\frac{1}{2}arctan\frac{U}{Q}$. 
The observation in 2019 has no polarization information so we show only total intensity (black line)
for this epoch in  Fig. \ref{fig:profiles}. For the other
epochs, we show linear polarization (red line), circular polarization (blue line), and the
measured PPAs in the upper part of the plots.
To obtain the uncertainty in the PPAs, we consider the unbiased linear polarization $L_{\rm true}$ 
as detailed in \cite{everett2001emission}. 
As scintillation
modifies the SNR from epoch to epoch, the number of well-defined PPA points varies accordingly.
The shown PPA values are corrected for Faraday rotation due to the ionospheric $\rm{RM}_{\rm ion}$ contribution 
by subtracting $\lambda^2 \times {\rm RM}_{\rm ion}$ from the PPA values (measured in radians), where $\lambda$ is the observing wavelength.

\section{Results}\label{sec:results}

\begin{figure}[htbp]
  \begin{center}
    \includegraphics[width= \columnwidth]{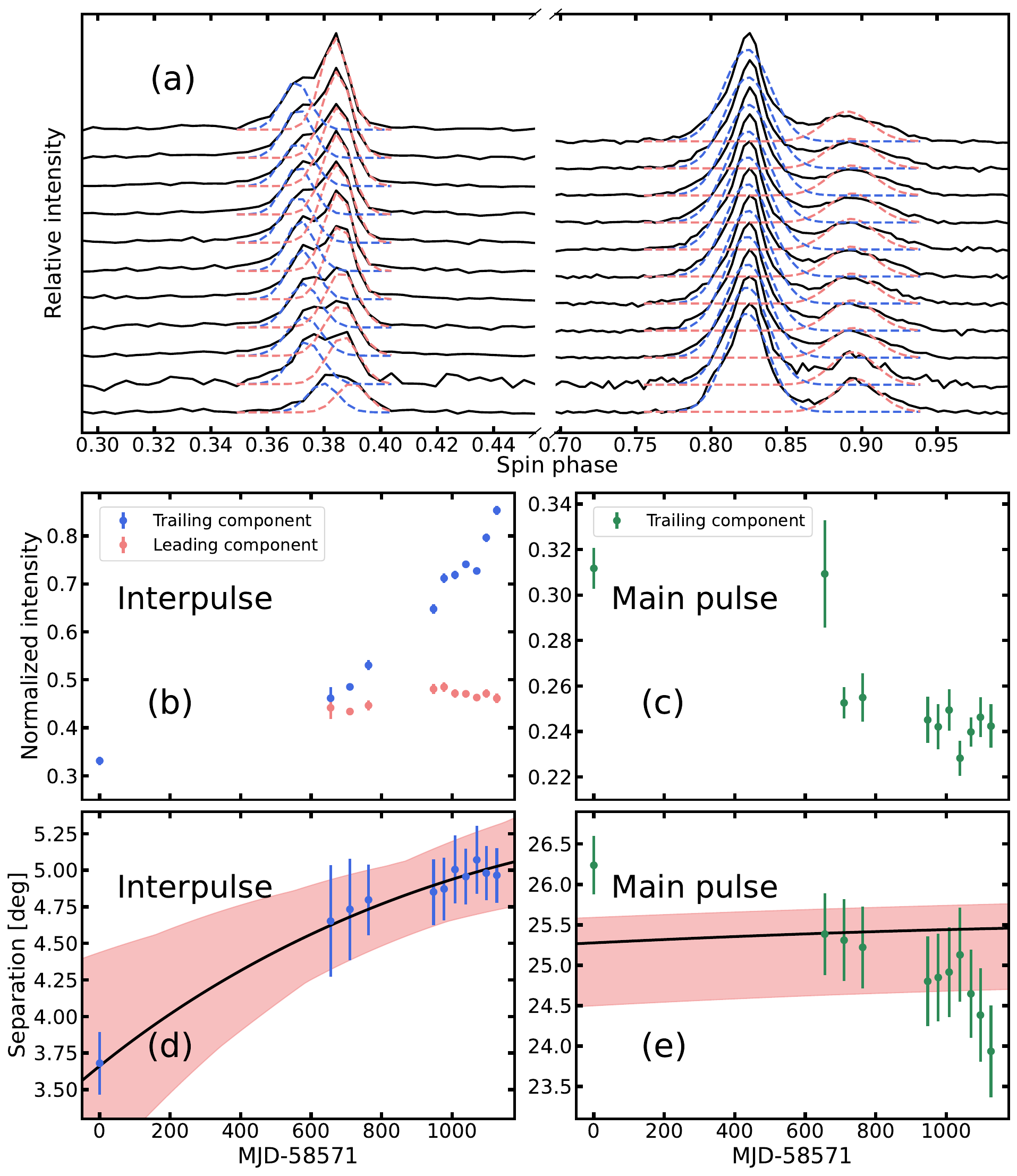}
\end{center}
\caption{We present the down-sampled pulse profile and the evolution in relative amplitude and separation of each pulse, providing insight into the temporal changes of PSR~J1946+2052's pulse profile. (a) The intensity profile of each epoch. The date of the observation goes from bottom to top. The dashed light coral and sea blue lines represent the Gaussian components that we derive from the fits of the pulse profiles.}  (b) The evolution of the interpulse's relative amplitude. (c) The evolution of the main pulse's relative amplitude. Only trailing components are shown because the leading components are used for normalization. (d) The evolution of the interpulse's separation between two components. (e) The evolution of the main pulse's separation between two components. The black solid and light coral lines in panels (d) and (e) indicate the best fit of our global fit model and samples in the $1-\sigma$ confidence interval. (see details in Sec. \ref{sec:sepfit}).\label{fig:zoominprofile}
\end{figure}

As shown in Fig. \ref{fig:profiles}, we have observed significant changes in the pulse profiles of PSR~J1946+2052 over a 3-year time span, which can be attributed to relativistic spin precession. 
The changes manifest themselves most prominently in a change in the 
relative amplitudes of both interpulse and main pulse over time.
To obtain a smooth pulse profile with less noise, we downsample the spin phase to 256 bins to extract the double-peak structure of both the interpulse and main pulse.
We normalize each integrated profile to the peak of the main pulse's leading component, and the relative amplitudes of other components are shown in panels (b) and (c) of Fig. \ref{fig:zoominprofile}.
The uncertainties in the relative amplitudes are calculated as the root mean squares of each profile's off-pulse region.

Relativistic spin precession can cause changes in the pulse profile and the PPAs.
To reveal these effects, we analyze the full polarization profile of PSR~J1946+2052.
We use a simple circular hollow-cone-shape emission beam to model the profile change.
For polarization, the PPA swings of PSR~J1946+2052 are not perfect S-shape, and the points shown in grey are not
easily consistent with a typical RVM swing. They may be caused by an unresolved orthogonal jump, as their location coincides with a minimum in linear polarisation and a change in the sign of circular polarisation. 
In our analysis, we then ignore this small range of PA values (displayed as
grey points in Fig. \ref{fig:profiles}). We present details of our analysis in the following sections.

\subsection{Pulse profile and PPA evolution}\label{sec:profile}

The most important profile evolution is that the interpulse has only one peak on 2019 March 30 but then
becomes double-peaked in observations taken after 2021 January 13.
Assuming a circular hollow-cone emission beam, this profile variation can be modelled by the movement of our line of sight from the edge of the emission beam to the interior of the hollow cone. 
This suggests that initially, the line of sight was near the edge of the beam, but our assumption of the beam shape
may be wrong \citep[cf. ][]{desvignes2019radio}.

\begin{figure}[htbp]
  \begin{center}
    \includegraphics[width= \columnwidth]{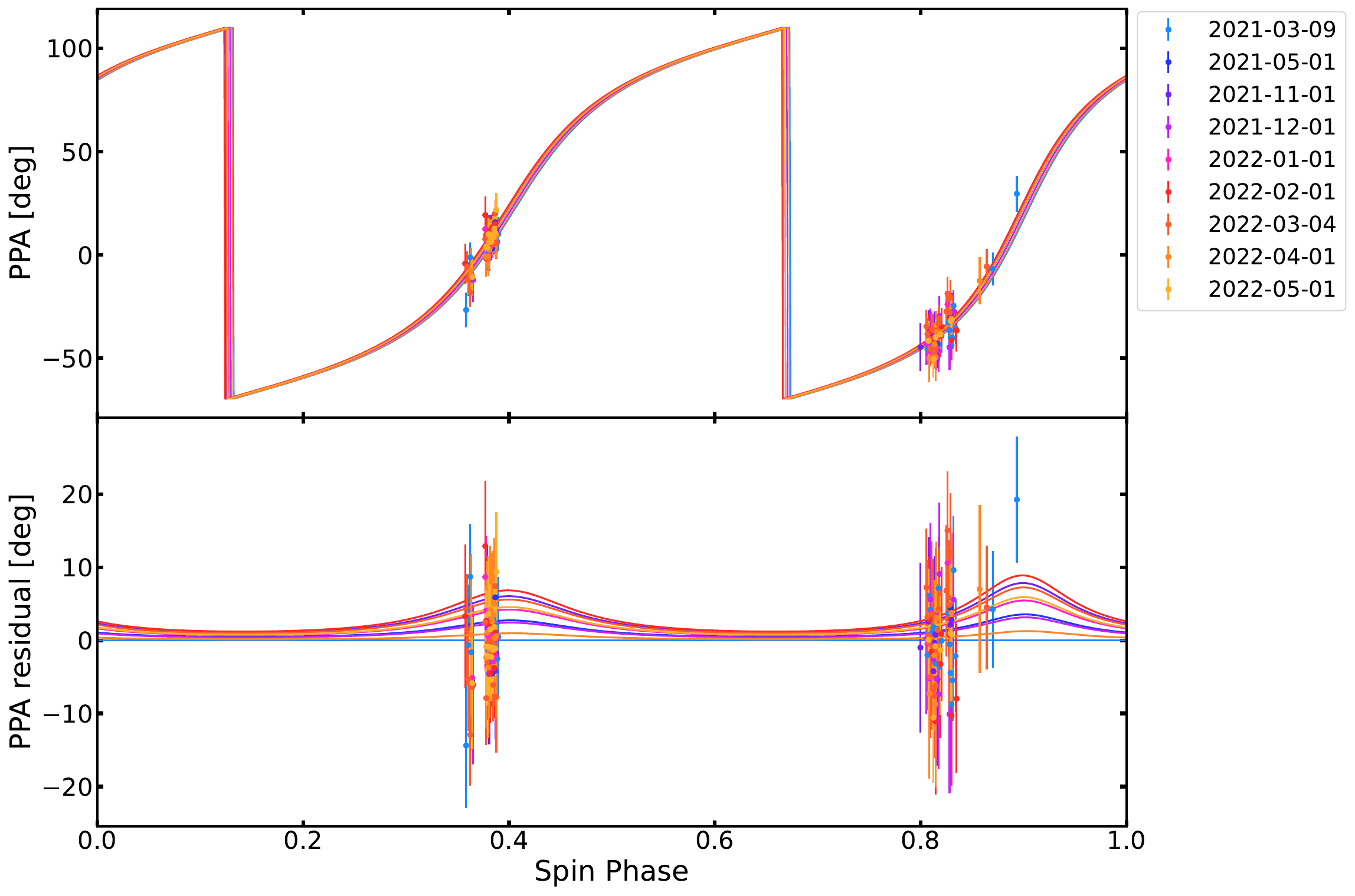}
\end{center}
\caption{In the upper panel, we display the PPA measurements at different epochs. The solid lines are the best-fit curves of the global fit model. In the lower panel, we display the residuals of the best-fit PPA swings based on the first curve on 2021-03-09 and also the residual between the PPA measurements and the best-fit PPA curves. We complete the points where the residual would jump for $180^\circ$.
\label{fig:PPA-bf}}
\end{figure}

To estimate the separations between the two components of the interpulse and the main pulse, we use two Gaussian functions with 
identical widths to fit these two pulse components separately in each observation.
The component separation is determined as the distance between the centres of the two Gaussians.
The separation between the interpulse and the main pulse does not enter the analysis at this stage.
For the interpulse seen on 2019-03-30 with only one peak, this method cannot be applied. 
Here, we fix the widths of the two Gaussian functions to the average value obtained from later observations. This fixed width is consistent with the distribution of the widths we derive from the 2021-2022 epochs. We also make the two Gaussian functions have the same amplitude.
With these additional constraints, we can obtain an estimated peak component separation for the 2019 observation. 
We display the Gaussian functions that are used to fit the interpulse and the main pulse in panel (a) of Fig. \ref{fig:zoominprofile}.
The peak component separations obtained using these methods are shown in panels (d) and (e) of Fig. \ref{fig:zoominprofile}.

Figure \ref{fig:PPA-bf} displays the stacked PPA measurements obtained from PSR~J1946+2052, which shows no clear evidence of evolution. 
The relatively large uncertainties and the absence of significant changes in the swing shape suggest that the expected spin precession has not yet reached a level of detectability through PPA measurements. This may be caused by the geometrical configuration, 
a special precession phase, or oversized uncertainties in our measurements.

\subsection{Geometry modeling}\label{sec:geofit}
\subsubsection{Convention and definitions}\label{sec:convention}

\begin{figure*}
  \begin{center}
    \includegraphics[width= 2\columnwidth]
    {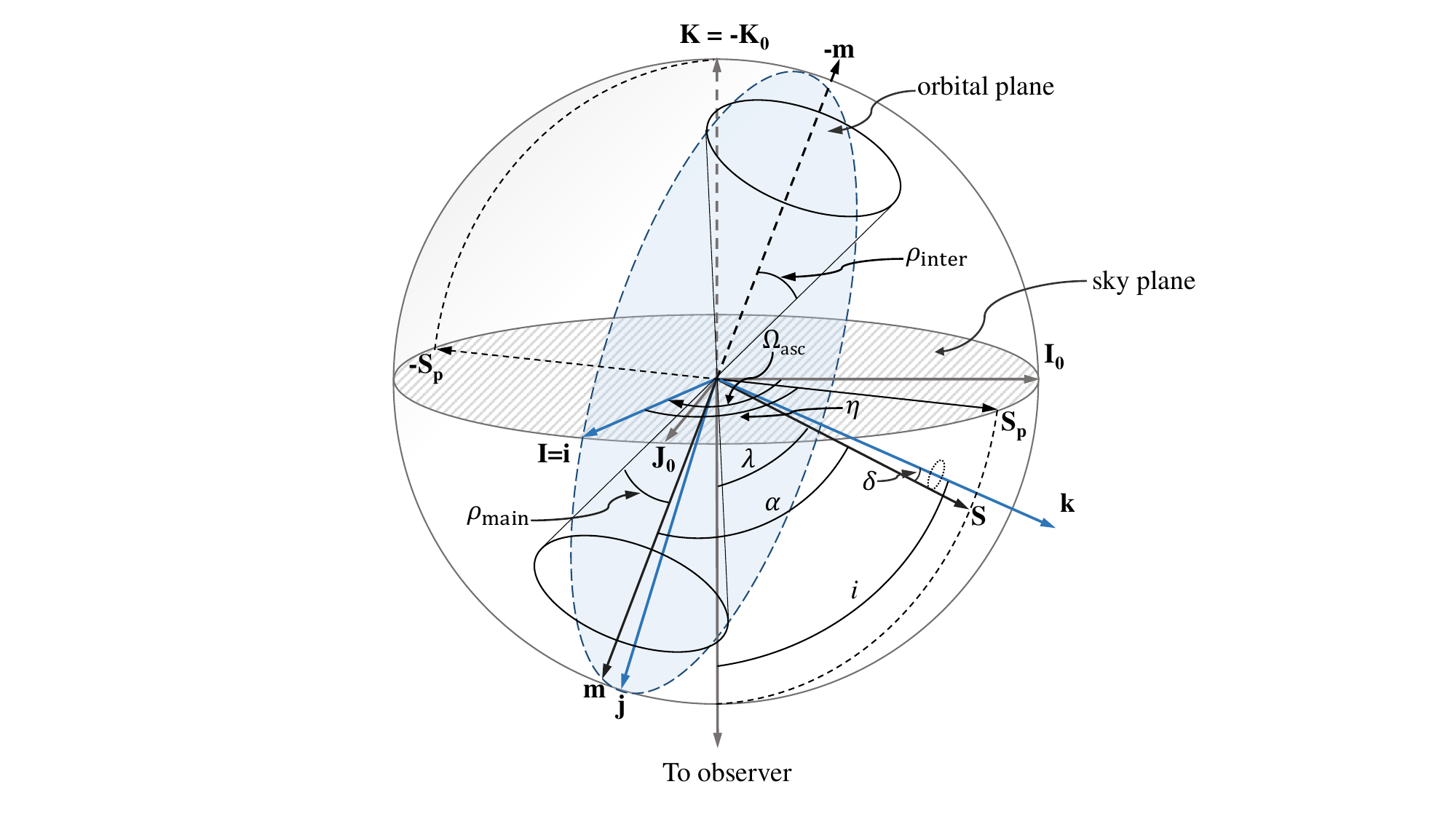}
\end{center}
\caption{We illustrate the geometrical angles in our convention. $\rm \mathbf I_0$ represents the North of the sky plane, and $\rm \mathbf J_0$ represents the East. The fundamental reference frame consists of $\rm \mathbf I_0$, $\rm \mathbf J_0$, and $\rm \mathbf K_0$. $\rm \mathbf S$ denotes the direction of the pulsar's spin vector, $\rm \mathbf S_p$ is the direction of the projection of the spin vector, and $\rm \mathbf k$ is the direction of the orbital angular momentum. The orbital reference frame consists of $\rm \mathbf i$, $\rm \mathbf j$, and $\rm \mathbf k$. The magnetic axis $\rm \mathbf m$, the main pulse's opening angle $\rho_{\rm main}$, and the interpulse's opening angle $\rho_{\rm inter}$ describe the emission beams. The definition of other angles can be found in the text.\label{fig:geo}}
 \end{figure*}

In this study, we adopt the PSR/IEEE convention \citep{2010PASA...27..104V} to define the geometrical angles and their relationships, which is commonly referred to as ``the observer's convention". 
These angles are illustrated in Fig. \ref{fig:geo}.
 It should be noted that this convention differs from the convention used in previous studies such as \cite{damour1992strong} and \cite{kramer2009double} (hereafter referred to as the DT convention) in which the line of sight is defined as pointing out from the sky plane. 
The angle between the line of sight and the spin vector ($\lambda$) and the inclination angle of the orbit ($i$) are complementary in these two conventions. Apart from those changes in convention, the formalism below
follows \cite{kramer2009double}.

As the spin angular momentum of the pulsar is much smaller than its total angular momentum, we can assume that the direction of the orbital angular momentum is constant with time and is the same as the total angular momentum. 
This means that the inclination angle $i$ remains constant with time.
As the pulsar's spin vector precesses around the orbital angular momentum, the line of sight will cut through different latitudes of the emission beam, causing the angle $\lambda$ to change with time according to the following function:
\begin{equation}
    {\rm cos}\,\lambda(t)={\rm cos}\,{\delta}\,{\rm cos}\,i+{\rm sin}\,{\delta}\,{\rm sin}\,i\,{\rm cos}\,\Phi_{\rm SO}(t),
\end{equation}
where $\delta$ is the misalignment angle of the spin vector from the orbital angular momentum vector. $\Phi_{\rm SO}(t)$ is the precession phase and is described as:
\begin{equation}
    \Phi_{\rm SO}(t)=\Phi_0 + \Omega_{\rm SO} (t-T_0),
\end{equation}
where $\Phi_0$ is the reference precession phase, $T_0$ is the time when the precession phase is $\Phi_0$ and $\Omega_{\rm SO}$ is the rate of the precession. In our work, we define $T_0$ as the date of the first observation (2019-03-30).

The projected position angle of the spin vector ($\Psi_0$) on the celestial sphere changes due to spin precession.
In our convention, the position angle equals 0 in the North of the sky plane ({$\rm \mathbf{I_0}$}) and increases counter-clockwise as turning to the East (\texttt{$\rm \mathbf{J_0}$}).  
$\Psi_0$ is described as
\begin{equation}
    \Psi_0(t)=\Omega_{\rm asc} + \Delta\psi_{\rm A} + \Delta_{\rm F} + \eta(t),
\end{equation}
where $\Omega_{\rm asc}$ is the ascending node of the orbital plane, $\Delta_{\rm A}$ is the angle shift caused by aberration, $\Delta_{\rm F}$ is the change of the Faraday rotation effect which is illustrated in Eq.~(\ref{eq:RM}), and $\eta(t)$ is the longitude of the pulsar's spin precession.
Because it is difficult to distinguish the specific values of $\Omega_{\rm asc}$, $\Delta\psi_{\rm A}$ and $\Delta_{\rm F}$, we simply define $\Psi_0(t)$ as
\begin{equation}
    \Psi_0(t)=\Delta\Psi + \eta(t),
\end{equation}
and $\eta(t)$ can be derived from
\begin{equation}
    {\rm cos}\,\eta(t)=\frac{{\rm sin}\,\delta\,{\rm sin}\,\Phi_{\rm SO}(t)}{{\rm sin}\,\lambda(t)},
\end{equation}

\begin{equation}
    {\rm sin}\,\eta(t)=\frac{{\rm cos}\,\lambda(t)\,{\rm cos}\,i-{\rm cos}\,\delta}{{\rm sin}\,i\,{\rm sin}\,\lambda(t)}.
\end{equation}

\subsubsection{Modeling pulse profile separations and PPAs}\label{sec:sepfit}
We can model the separation evolution by assuming a circular hollow-cone-shape emission beam.
For the dipole field emission, the separation between the interpulse and the main pulse could deviate from $180^\circ$ due to the different emission heights, while the shape of these two pulses could be the hollow-cone shape.
We assume each hollow-cone beam to be symmetrical so that the separation between two peak components in one pulse can be expressed as \citep{kramer1998determination}:
\begin{equation}
    {\rm sin}^2 \left(\frac{{\rm W}(t)}{4}\right)=\frac{{\rm sin}^2\,(\rho/2)-{\rm sin}\,^2(\beta(t)/2)}{{\rm sin}\,{\alpha}\,{\rm sin}\,(\alpha+\beta(t))},
\end{equation}
where $\rm W(t)$ is the separation, $\rho$ is the opening angle of the emission beam, $\alpha$ is the magnetic inclination angle and $\beta(t)$ is the angle between the magnetic axis and the line of sight. 
$\beta(t)$ is given by:
\begin{equation}
    \beta(t)=\lambda(t)-\alpha.
\end{equation}
Once we determine $\alpha$ for the main pulse, the interpulse's magnetic inclination angle is given by $\pi-\alpha$.
We have observed the interpulse changing from a one-peak to a two-peak shape,
but the short time span covered by our observations 
alone is insufficient to constrain the six geometrical parameters that we need to model the precession.

For modelling the PPAs, we continue to
adopt the precessional RVM \citep{kramer2009double} that considers relativistic spin precession. 
The conventional RVM is used to model the PPA that increases clockwise. However, the PPA we use increases counter-clockwise as we derive the Stokes parameters from \texttt{PSRCHIVE}, which follows the PSR/IEEE convention \citep{2010PASA...27..104V}. 
To modify the RVM equation accordingly, we refer to \cite{everett2001emission}.
As $\Psi_0(t)$ and $\beta(t)$ vary with time due to precession, the precessional RVM can be described by Eq.~(\ref{eq:rvm}),
\begin{equation}
\begin{aligned}
            &\Psi(t)  = \Psi_0(t) + \nonumber
            \\&{\rm tan^{-1}}\left(\frac{{\rm sin}\,{\alpha}\,{\rm sin}\,(\phi_0-\phi)}{{\rm sin}\,{\lambda(t)}\,{\rm cos}\,{\alpha}-{ \rm cos}\,{\lambda(t)}\,{\rm sin}\,{\alpha}\,{\rm cos}\,(\phi_0-\phi)}\right),
        \label{eq:rvm}
\end{aligned}
\end{equation}
where $\phi$ is the spin phase, $\phi_0$ is the reference phase of the spin axis and is relative to the main pulse.
Due to the limited SNR of our polarization data, we cannot consider the difference in emission height between the interpulse and the main pulse through PPAs, as done in \cite{2019MNRAS.490.4565J}.
Although the reference phase of the spin axis position angle is not expected to change due to relativistic spin precession, we fit $\phi_0$ for every epoch because the profiles are aligned based on the maximum of the main pulse.
We exclude the part of the PPA where apparent orthogonal polarization modes are present. 
The selected PPA points are shown in Fig. \ref{fig:profiles} as black points.

In order to constrain the geometry using profile separations and PPAs, we combine the profile separation model and precessional RVM and reconstruct the likelihood as given by Eq.~(\ref{eq:likelyhood}).
\begin{equation}
\begin{aligned}
    {\rm ln\mathcal{L}}=
    &-\frac{1}{2}\sum\limits_{\rm i=1}^{ N_{\rm MJD}}\left(
    \sum\limits_{\rm j=1}\limits^{N_{\rm PPA,i}}C_{\rm PPA,i}\left(\frac{\Psi^{\rm obs}-\Psi}{\sigma_{\rm PPA}^{\rm obs}}\right)_{\rm j}^2\right.\nonumber
    \\&+C_{\rm inter}\left(\frac{\rm W_{inter}^{obs}-W_{inter}}{\sigma_{\rm inter}^{\rm obs}}\right)^2\nonumber
    \\&\left.+C_{\rm main}\left(\frac{\rm W_{main}^{\rm obs}-W_{\rm main}}{\sigma_{\rm main}^{\rm obs}}\right)^2\right)_{\rm i},
    \label{eq:likelyhood}
\end{aligned}
\end{equation}
% \end{figure}
where i and j indicate each day and each PPA point in day i's observation.
As shown in Eq.~\ref{eq:likelyhood}, we weigh each component of the likelihood with the number of data points as follows:
\begin{equation}
 C_{\rm inter}=C_{\rm main}=1,\ C_{\rm{PPA},i}=\frac{N_{\rm inter}}{N_{\rm{PPA},i}},
\end{equation}
where $C_{\rm inter}$, $C_{\rm main}$ and $C_{\rm PPA,i}$ are the weighted coefficients we used for the component separations of the interpulse, the main pulse and each day's PPA swing. $N_{\rm inter}$ is the number of data points of the interpulse's separation. Note that $N_{\rm inter}=N_{\rm main}$ because we obtain one data point for the component separation of the main pulse and interpulse in each observation. $N_{\rm PPA,i}$ indicates the number of data points of each day's PPA swing.

This likelihood enables us to fit the profile separations with RVM information of $\beta$ and ensures that $\beta<\rho$. 
% This likelihood can provide the fit of profile separations with RVM information of $\beta$ and make sure $\beta<\rho$.
We utilize the Markov Chain Monte Carlo (MCMC) method based on the Python packages \texttt{emcee} \citep{emcee} and \texttt{corner} \citep{corner} to obtain the distribution of geometrical parameters. 

Since we have no prior information on the geometry, we search for the optimal geometry parameters with uniform
priors over the allowed range while assuming that GR is correct. 
The individual masses of the binary system are not accurately determined yet. Still, we can
use the mass function ($0.268184(12)~M_{\odot}$) and total mass ($2.50(4)~M_{\odot}$) of this system from \cite{stovall2018palfa}, 
to express the precession rate as a function of the orbital inclination angle $i$.

\begin{figure*}
  \begin{center}
    \includegraphics[width= 2\columnwidth]{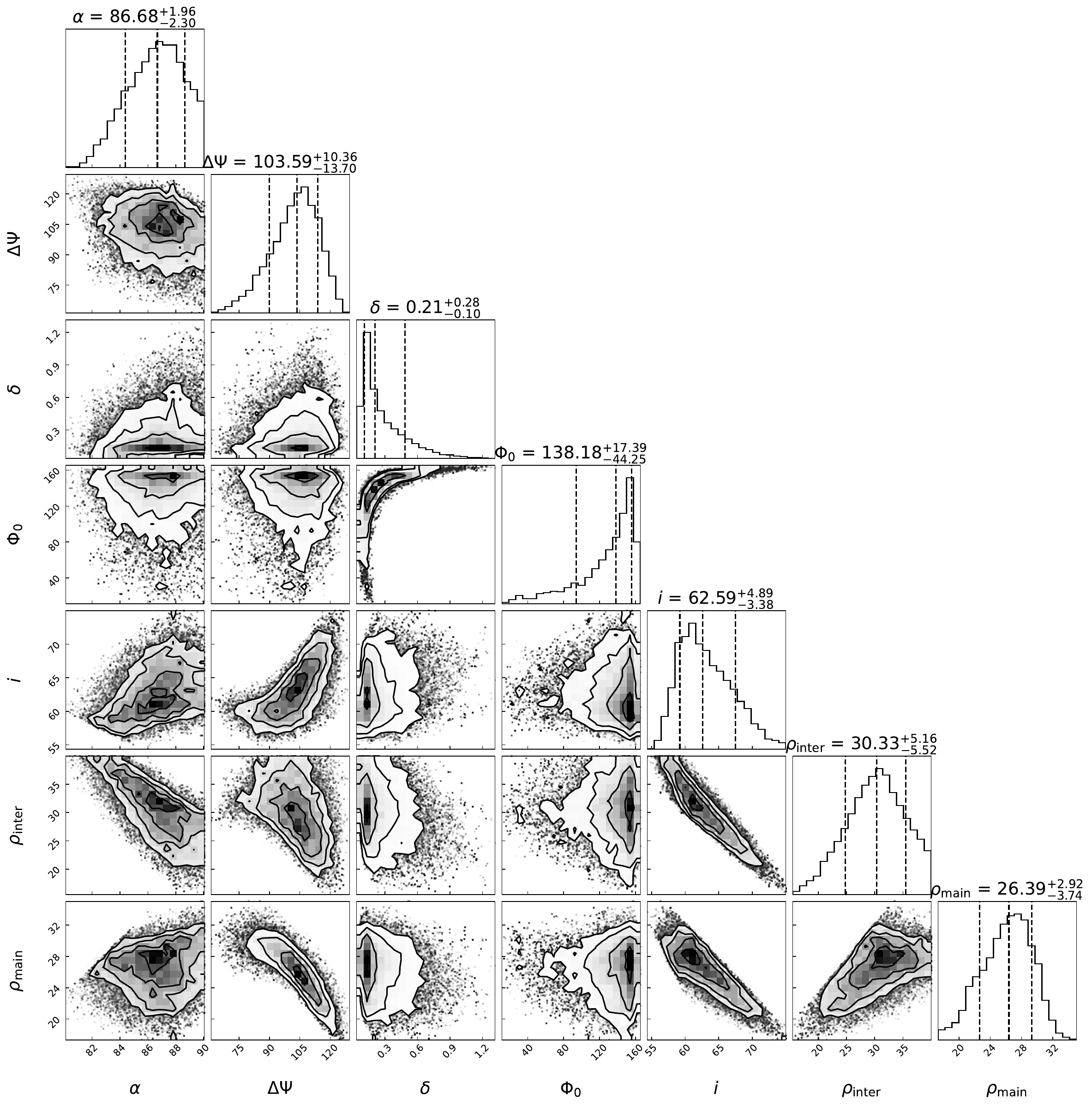}
\end{center}
\caption{The posterior distribution of the geometrical parameters in the global fit. The angles are given in units of degree. The value of each parameter indicates the result in a 68\% confidence level.}  \label{fig:contour}
\end{figure*}

\begin{figure*}
  \begin{center}
    \includegraphics[width= 2\columnwidth]{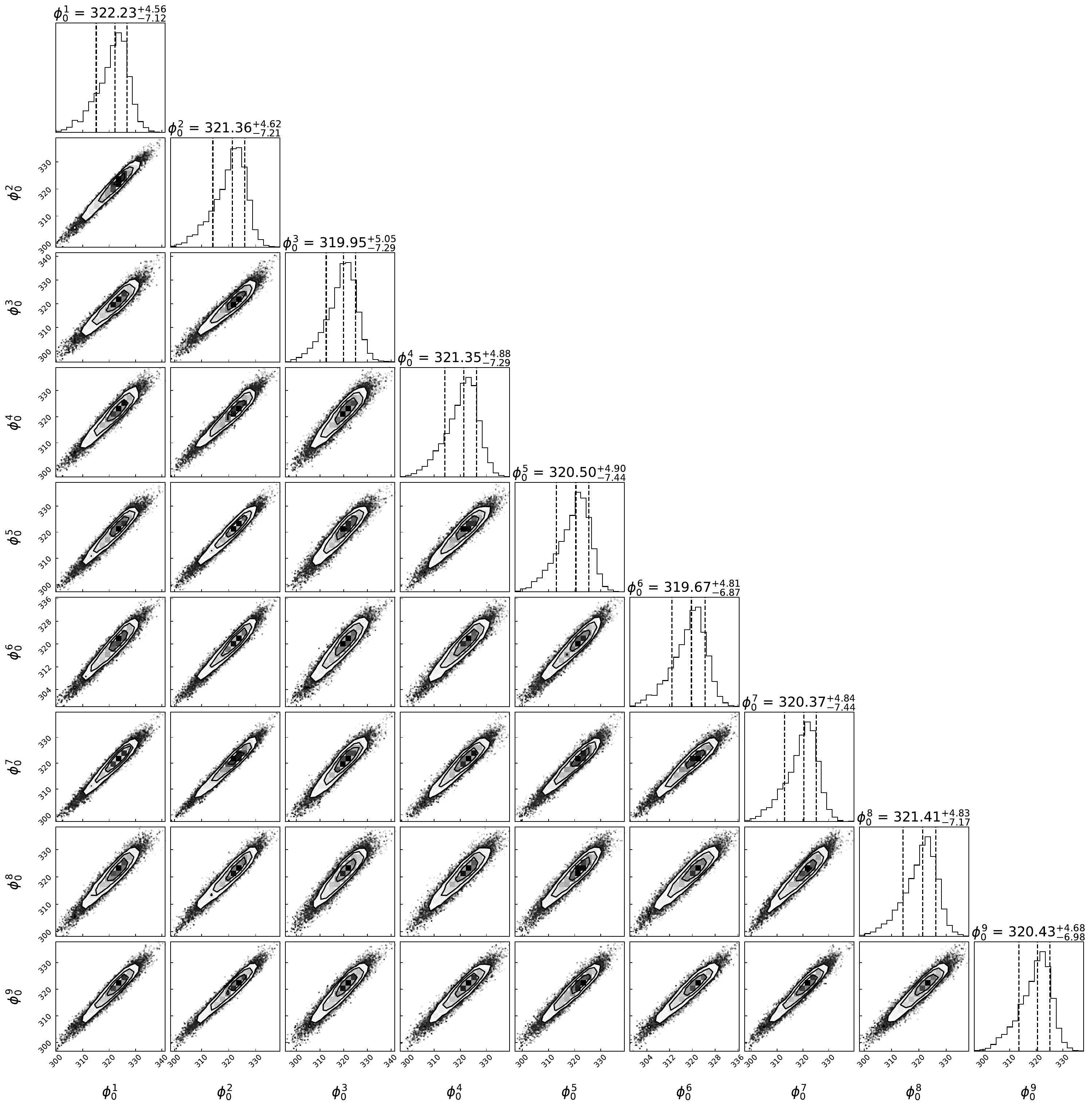}
\end{center}
\caption{The posterior distribution of the $\phi_0$ in the global fit. The superscripts 1 to 9 represent the date from 2021-03-09 to 2022-05-01. The angles are given in units of degree. The value of each parameter indicates the result in a 68\% confidence level.}\label{fig:contour2}
\end{figure*}

\begin{figure}[htbp]
  \begin{center}
    \includegraphics[width= 0.8\columnwidth]{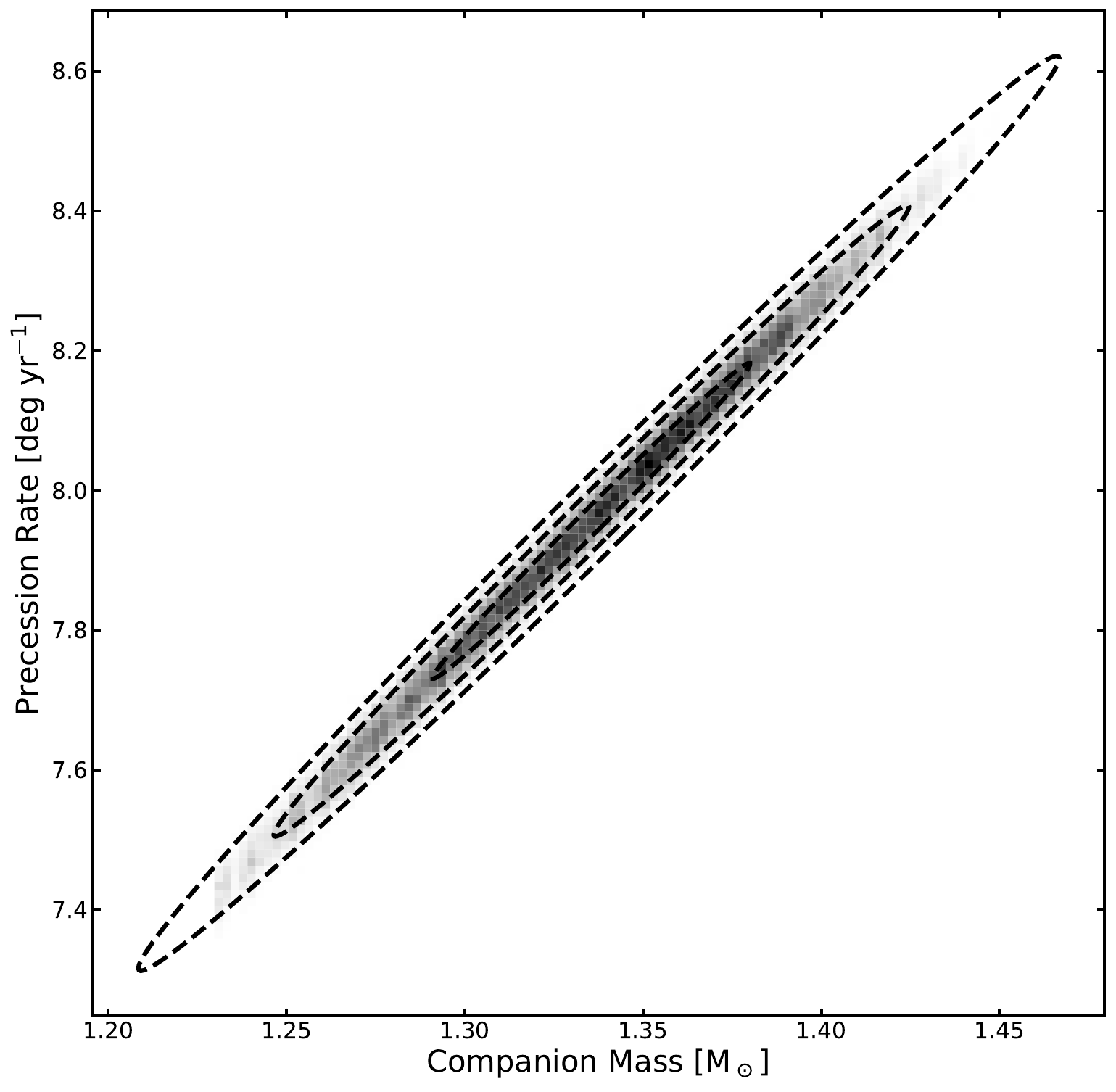}
\end{center}
\caption{The distribution of the precession rate and companion mass. We display the confidence levels of 1-$\sigma$, 2-$\sigma$, and 3-$\sigma$ with dashed lines.\label{fig:OmegaM2}}
\end{figure}

\begin{figure}[htbp]
  \begin{center}
    \includegraphics[width= 1\columnwidth]{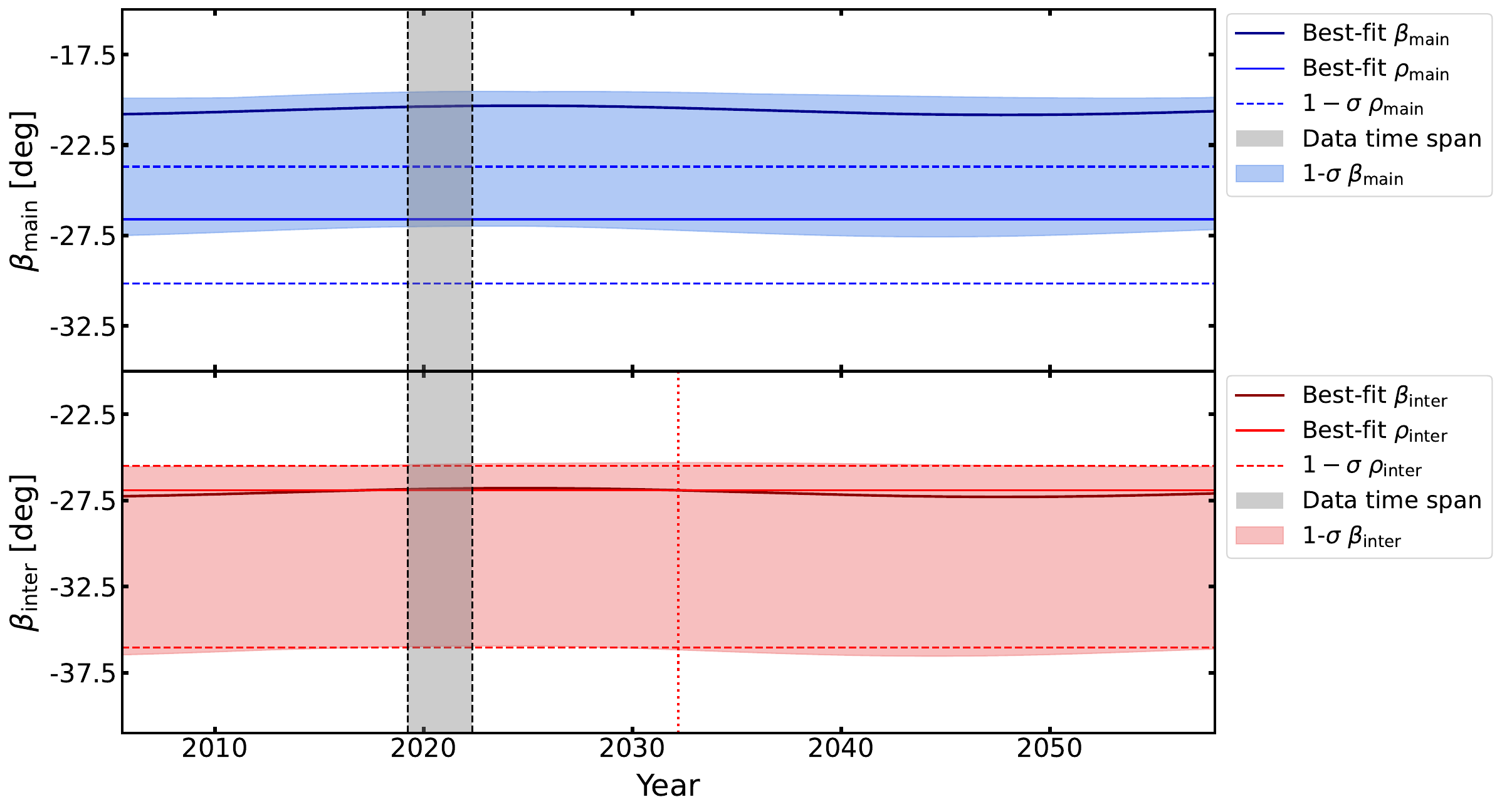}
\end{center}
\caption{The evolution of $\beta$ derived from the best fit geometrical parameters of PSR~J1946+2052. For the upper panel, the blue solid and dashed lines indicate the best fit $\rho$ of the main pulse and its 1-$\sigma$ interval. The dark blue and translucent blue curves indicate the best fit $\beta$ of the main pulse and samples in the 1-$\sigma$ interval. For the lower panel, the red solid and dashed lines indicate the best fit $\rho$ of the interpulse and its 1-$\sigma$ interval. The dark red and translucent red curves indicate the best fit $\beta$ of the interpulse and samples in the 1-$\sigma$ interval.
The grey area is the time span of our observations.
The red dotted line in the lower panel indicates the day when the separation of the interpulse disappears, which is predicted to be around 2032.\label{fig:betachange}
}
\end{figure}

We display the MCMC result of geometrical angles in Fig. \ref{fig:contour} and $\phi_0$ of each epoch in Fig. \ref{fig:contour2}.
We present in Fig.~\ref{fig:PPA-bf} the best fit of the PPA swings, while the best fit of separations and 1-$\sigma$ samples are displayed in Fig.~\ref{fig:zoominprofile}.
% We present the best-fit and the 1-$\sigma$ line of the PPA swings fit in Fig. \ref{fig:PPA-bf}.
% The best-fit of separations are displayed in Fig. \ref{fig:zoominprofile}.
Our results show that $\delta$ is very small, and is consistent with the expectation by \cite{stovall2018palfa}. 
The value of $\alpha$ is related to the main pulse and indicates that the pulsar is a nearly orthogonal rotator.
The observed significant difference in profile separations between the interpulse and the main pulse is because the spin axis and the magnetic axis are not entirely orthogonal. 
Given the small value of $\delta$, we cannot derive a precise precession phase $\Phi_0$.
Assuming negligible effects from additional Faraday rotation and aberration, the longitude of the ascending node $\Omega$ can be determined to a small range.
The inclination angle $i$ has been constrained by the fit to ${i=63^\circ}^{+5^\circ}_{-3^\circ}$, which is consistent with the 
updated timing result (Freire, private communication) where $i$ is estimated to be between $65^\circ$ and $80^\circ$.
Based on our estimate of $i$, we can predict the companion mass and the precession rate to be $1.33(4)~ \rm{M_\odot}$ and $7.96(23)^\circ\ {\rm yr}^{-1}$ respectively. 
Thus the mass of PSR J1946+2052 is predicted to be $1.17(6)~ \rm{M_\odot}$.
The distribution of these two parameters is shown in Fig. \ref{fig:OmegaM2}.
By calculating $\beta$ with the best fit geometrical parameters (the dark red curve in Fig. \ref{fig:betachange}), we predict that the separation of the interpulse will disappear in 2032 because our line of sight will move out of the hollow-cone shape emission beam.
While due to the small amplitude of $\beta$, the separation of the main pulse will stay visible all the time.
We display the evolution of $\beta$ in Fig. \ref{fig:betachange}.
With improved data and longer time baselines, in the future, we may be able to combine the timing analysis with profile and polarization studies to derive the geometry of this system precisely and even test GR.

\section{Discussion and Conclusion}\label{sec:discussion}

Through our analysis of the profile separations and PPAs, we are able to determine that the spin-orbit misalignment angle $\delta$ of the PSR J1946+2052 system is very small, approximately $0.2^\circ$ (as shown in Fig. \ref{fig:contour}). 
Due to the pulsar's short spin period, it appears to be a recycled pulsar and therefore the primary neutron star in the system. 
The small value of $\delta$ suggests the spin axis and the orbital angular momentum to be nearly aligned, consistent with the hypothesis that the kick from the second supernova (SN) was small \citep{stovall2018palfa}.
It is worth noting that another system with a small $\delta$ ($<3.2^\circ$) is the recycled pulsar 
in the double pulsar system, PSR J0737$-$3039A \citep{ferdman2013double}.
The small $\delta$ and low eccentricity of both PSR~J0737$-$3039A ($e=0.088$) and PSR~J1946+2052 ($e=0.064$) imply that the second SN in each system was likely low-kick and symmetric, with little mass loss.
In the case of the double pulsar, accurate mass measurements of both pulsars through pulsar timing have provided additional insight into the nature of the second SN.
The low mass of the companion pulsar PSR J0737$-$3039B and the low system velocity suggests that the second SN in this system could be an ultra-stripped SN \citep{2013ApJ...778L..23T,tauris2017formation} or an electron-capture SN \citep{ferdman2013double}.
Based on the companion mass we derive, we infer that the second SN of the PSR~J1946+2052 system is likely similar to that of the double pulsar system.
\cite{ferdman2013double} did not find obvious profile evolution of PSR~J0737$-$3039A from 2005 to 2011, confirmed more recently by \cite{2022A&A...667A.149H}.
The reason why we nevertheless observe profile evolution in PSR~J1946+2052 is likely that our line of sight is closer to the edge of the circular emission beam than that in PSR~J0737-3039A, thus the variation of the peak separation is significant. 

The trend of the separation change is different for the interpulse and the main pulse.
The separation of the interpulse is increasing, while that of the main pulse appears to be decreasing.
The variation in the interpulse from one peak to two peaks suggests that our line of sight is moving towards 
the magnetic axis of the interpulse if the beam is a hollow cone.
In this work, we make the assumption that the two radio beams of the pulsar are symmetrical. 
Based on our fitting result, the line of sight is near the edge of two emission beams and moving toward the centre of the beams.
Therefore we would expect the separation between the main pulse peaks to increase as well, which is, however, the opposite.
This could be due to the complexity of the emission pattern in the interior of the main pulse's emission beam, as shown in PSR~J1906+0746 \citep{desvignes2019radio}.

In this work, we have assumed the validity of GR and used a simple beam model to constrain binary parameters.
To determine when we can utilize PSR~J1946+2052 for testing GR with spin precession, we conduct simulations of future observations of profile separations and PPA swings, assuming that our current best-estimated geometry is accurate.
We produce a set of simulated profile peak separations and PPA swings every two months, based on the projections from our best-fit parameters, and incorporate the precession rate we derive from these parameters.
To account for measurement errors on the actual profile peak separations and position angle swings, we introduce noise to our projected data.
The error of each data point is determined by using the bootstrap method among corresponding existing errors.
As the number of new simulated PPA swings grows faster than the number of simulated profile component separations, PPA swings will dominate the global fit.
To counterbalance this discrepancy, we resample the simulated PPA swings and average the values for each year. This should not impact our fit outcome, as the position angle swing varies very slowly with time.

We perform simulations for additional 3-year, 6-year, and 9-year data sets, and present the results in Table \ref{tab:simutab}, except $\phi_0$ of each epoch.
We fit for the value of $\Omega_{\rm SO}$ without assuming GR and find that it can be constrained to a relative precision of only 43\% in 9 years (total of 11 years).
However, these simulations do not incorporate pulsar timing. 
With 11-year FAST timing, we could possibly derive the DNS's masses and inclination angle $i$ as well as potentially the system's longitude angle of the ascending node $\Omega_{\rm asc}$.
The timing measurement could be used as the prior in our analysis of the profile and PPA, significantly improving the constraint on the other geometrical parameters and the spin precession rate under the assumption of GR.
This improvement is important for the pulsar emission model and binary evolution.

In summary, we observed
PSR~J1946+2052 for 11 times and find significant profile evolution and attribute this evolution to relativistic spin precession.
We use the circular hollow-cone emission beam to derive the geometry of this system.
We derive a global fit model by combining the separation fit and PPA fit to estimate the geometrical angles.
In the fit, we follow the results of PSR~J1946+2052's mass function and total mass in \cite{stovall2018palfa} and assume GR to be correct.
The result shows that the spin vector of this pulsar is nearly aligned with the orbital angular momentum, indicating that the second SN should be fast and symmetric, as suggested by \citet{stovall2018palfa}. 
We also constrain the companion mass and the precession rate under the scenario of GR. We emphasize that the
derived uncertainties do not take into account possible systematic shortcomings of the assumed beam model.
The future timing of PSR J1946+2052 will improve our understanding of the pulsar emission model as well as binary evolution and constrain the spin precession rate under the assumption of GR.

\begin{table*}[t]
\centering
\caption{Geometrical parameters in simulations}
\label{tab:simutab}
 \begin{tabular}{c c c c c c c c c} 
 \hline
 Data & $\alpha$ & $\Delta\Psi$ & $\delta$ & $\Phi_0$ & $i$ & $\rho_{\rm inter}$ & $\rho_{\rm main}$ & $\mathbf{\Omega_{\rm SO}}$ \\ 
      & (deg)  & (deg) & (deg) & (deg) & (deg) & (deg) & (deg) & (deg ${\rm yr}^{-1}$)\\[0.5ex]
 \hline

  3-year & $86.81^{+0.28}_{-0.29}$ & $110\pm1$ & $0.35^{+0.85}_{-0.24}$ & $144^{+17}_{-31}$ & $66\pm1$ & $27\pm1$ & $24\pm1$  & $\mathbf{6^{+5}_{-3}}$  \\ [1ex]

  6-year & $86.78\pm0.15$ & $108\pm1$ & $0.45^{+0.83}_{-0.31}$ & $150^{+12}_{-26}$ & $65\pm1$ & $27.78\pm0.45$ & $24.53\pm0.29$ & $\mathbf{5^{+5}_{-2}}$  \\ [1ex]
 
  9-year & $86.68^{+0.10}_{-0.11}$ & $112\pm1$ & $0.30^{+0.38}_{-0.15}$ & $142^{+13}_{-18}$ & $66.43^{+0.39}_{-0.45}$ & $26.64\pm0.40$ & $23.39^{+0.26}_{-0.27}$ & $\mathbf{7^{+3}_{-2}}$ \\ [1ex]

 \hline
\end{tabular}
\end{table*}

\section{Acknowledgements}
We appreciate helpful comments and discussions with Emmanuel Fonseca. This work was supported by the National SKA Program of China (2020SKA0120200, 2020SKA0120300), 
the National Natural Science Foundation of China (12203072, 12041303, 12203070), and the CAS-MPG LEGACY project. 
Lijing Shao is supported by the Max Planck Partner Group Program funded by the Max Planck Society.
Xueli Miao and Mengyao Xue are supported by the Cultivation Project for FAST Scientific Payoff and Research Achievement of CAMS-CAS.
Jumei Yao is supported by the Natural Science Foundation of Xinjiang Uygur Autonomous Region (grant No. 2022D01D85) and the Major Science and Technology Program of Xinjiang Uygur Au320.50tonomous Region (grant No. 2022A03013-2).

\bibliography{main}{}
\bibliographystyle{aasjournal}

\end{sloppypar}
\end{document}